\documentclass[a4paper,12pt]{article}
\usepackage{graphicx,epsfig}
\usepackage{epstopdf}
\usepackage{amssymb}
\usepackage{amsmath}

\newcommand{\mrm}[1]{\mbox{\rm #1}}

\newcommand{\Eq}[1]{Eq.(\ref{#1})}

\newcommand{\be}{\begin{equation}}
\newcommand{\ee}{\end{equation}}
\newcommand{\br}{\begin{eqnarray}}
\newcommand{\bea}{\begin{eqnarray}}
\newcommand{\eea}{\end{eqnarray}}
\newcommand{\er}{\end{eqnarray}}
\newcommand{\ba}{\begin{array}}
\newcommand{\ea}{\end{array}}
\newcommand{\bi}{\begin{itemize}}
\newcommand{\ei}{\end{itemize}}
\newcommand{\bn}{\begin{enumerate}}
\newcommand{\en}{\end{enumerate}}
\newcommand{\bc}{\begin{center}}
\newcommand{\ec}{\end{center}}

\def\gappeq{\mathrel{\rlap {\raise.5ex\hbox{$>$}}
{\lower.5ex\hbox{$\sim$}}}}

\def\lappeq{\mathrel{\rlap{\raise.5ex\hbox{$<$}}
{\lower.5ex\hbox{$\sim$}}}}

\begin{document}
\pagestyle{empty}
\begin{flushright}
\end{flushright}
\vspace*{25mm}
\begin{center}
{\large {\bf Direct determination of neutrino mass parameters at future colliders}} \\
\vspace*{2cm}
{\bf M. Kadastik},  {\bf M. Raidal} and {\bf L. Rebane}
\vspace{0.3cm}

National Institute of Chemical Physics and Biophysics, Ravala 10,
Tallinn 10143, Estonia \\

\vspace*{3cm}
{\bf ABSTRACT} \\ 
\end{center}
\vspace*{5mm}
\noindent
If the observed light neutrino masses are induced by their Yukawa couplings to singlet 
right-handed neutrinos, natural smallness of those renders direct collider tests of 
the electroweak scale neutrino mass mechanisms almost impossible both in the case of  
Dirac and Majorana (seesaw of type I) neutrinos. However, in the triplet Higgs seesaw
scenario the smallness of light neutrino masses may come from  the smallness of $B-L$
breaking parameters, allowing sizable Yukawa couplings even for a TeV scale triplet. We show
that, in this scenario, measuring the branching fractions of doubly charged Higgs to 
different same-charged lepton flavours at LHC and/or  ILC experiments will allow  one 
to measure the neutrino mass parameters which neutrino oscillation experiments are insensitive to,
 including the neutrino mass hierarchy, lightest neutrino mass and Majorana phases.
\vspace*{3cm}
\noindent

\begin{flushleft} 
January  2008
\end{flushleft}
\vfill\eject

\setcounter{page}{1}
\pagestyle{plain}

\section{Introduction}

 In recent past neutrino oscillation experiments have shown convincingly that at least two 
  light neutrinos have  
 non-zero masses and their mixing is characterized by two large mixing 
 angles \cite{GonzalezGarcia:2002dz}. 
 Those facts constitute
 indisputable evidence of new physics beyond the standard model (SM). 
However, despite of the intense experimental and theoretical effort over many years,  
 understanding of the origin of neutrino masses is still missing. 

From the experimental point of view the information on neutrino masses coming from oscillation 
experiments is limited by the fact that these experiments are only able to measure the differences of squared neutrino masses and not their absolute magnitude, neither are they sensitive to the Dirac or
Majorana nature of light neutrinos.
In particular, the present oscillation experiments cannot distinguish between the two possible mass ordering patterns of light neutrinos, the normal and the inverted ones, and are insensitive to the possible Majorana phases \cite{81}  of neutrinos. 
The observed smallness of 
neutrino $\theta_{13}$ mixing angle makes it very difficult to measure any new parameter, such as the 
neutrino Dirac CP phase $\delta,$ in neutrino oscillation experiments before a distant-future neutrino factory \cite{Huber:2004ug}. 
To learn conceptually new  facts about light neutrinos in a shorter time-scale requires  likely
an experimental breakthrough either in low energy neutrino
experiments, such as the neutrinoless double beta decay ($0\nu\beta\beta$) decay experiments, or in collider physics.

 From the theory side we still do not know why neutrinos are so light compared to charged fermions.
 It is natural that the  $SU(2)_L$ doublet neutrinos couple to 
new singlet (right-handed) neutrinos $N$ and the SM Higgs doublet in a direct analogy with
 all other SM fermions. If this is the only new physics, the smallness of neutrino masses 
 requires unnaturally small Dirac Yukawa couplings.
 Alternatively, the singlet neutrinos  may have very large Majorana masses which suppress  the light
 neutrino masses to the observed range via the seesaw mechanism of type I \cite{Minkowski:1977sc,Yanagida:1979as,Gell-Mann:1980vs,Glashow:1979nm,Mohapatra:1979ia}
 even for large values of the Yukawa couplings.  Generically neither of those simple scenarios
 can be directly probed at low energy nor collider experiments\footnote{This conclusion may be different if softly broken supersymmetry exists in Nature.  Flavour violating Yukawa couplings of heavy
 neutrinos may induce flavour off-diagonal elements in the soft slepton mass matrices via the renormalization effects \cite{Borzumati:1986qx,Hall:1985dx}  which may lead to observable rates of lepton flavour violating processes.
 This very complex scenario requires analyses beyond the present one \cite{Raidal:2008jk}.}. 
 Making the singlet neutrinos as light as 1 TeV to be kinematically accessible at colliders  does not 
 help because their only interactions are of Yukawa type and the seesaw mechanism predicts that
 the couplings are too small for any observable signal except the neutrino masses.  Complicated
 model building is required to ensure the correct light neutrino masses, 1 TeV heavy neutrinos
 and meaningfully large neutrino Yukawa couplings at the same time. Unfortunately the direct 
 tests of singlet neutrino mass mechanism at LHC 
 are experimentally demanding even in those models \cite{Han:2006ip,Bray:2007ru,Aguila:2007em}.

However, group theory tells us that generation of non-zero masses for the SM doublet neutrinos 
does not require the existence of singlets. One of the best motivated and best studied
neutrino mass scenario is the triplet Higgs mechanism \cite{Magg:1980ut,Schechter:1980gr,Lazarides:1980nt,Mohapatra:1980yp,Gelmini:1980re}, 
sometimes called seesaw mechanism of type II.
From the point of view of direct tests the triplet neutrino mass 
mechanism has several advantages over 
the singlet one. Firstly, the $SU(2)_L$ triplet multiplet  contains a doubly charged scalar 
which can be pair produced at colliders independently of their Yukawa couplings. 
Thus tests of this mechanism are limited only by the collision energy. Secondly, the smallness of
neutrino masses  does not imply the smallness of triplet Yukawa couplings. As neutrino masses
in this scenario are necessarily of Majorana type, they may be different from the Dirac
fermion masses because of the smallness of $B-L$ breaking. This is natural by the 't Hooft
criterion as $B-L$ is a conserved quantum number in the SM. Thus the neutrino Yukawa couplings
to triplet may be  sizable, constrained by unobserved lepton flavour violating interactions,
and dominate over the triplet coupling to two gauge bosons. Thirdly, the triplet Yukawa couplings 
directly induce the neutrino mass matrix up to the small  $B-L$ breaking triplet vacuum expectation
value (VEV) which appears in neutrino masses  as a common proportionality factor.
Altogether those arguments imply that one can  study the neutrino mass parameters
at Large Hadron Collides (LHC) and/or International Linear Collider (ILC)
 experiments by just counting flavours of the same-charged lepton pairs 
originating from the doubly charged Higgs boson decays.

 In this work we extend the analyses of our previous paper \cite{Hektor:2007uu}.
 While in Ref. \cite{Hektor:2007uu} we studied the discovery potential of LHC experiments 
 for the process $pp\to\Phi^{++}\Phi^{--} $~\cite{Gunion:1989in,Huitu:1996su,Gunion:1996pq,Dion:1998pw,Muhlleitner:2003me,Akeroyd:2005gt,Hektor:2007uu,Han:2007bk}  
 assuming that the subsequent decays of $\Phi^{\pm\pm} $
 are determined by neutrino data, in this analyses we turn the argument around and study
 what can one learn about neutrino physics if LHC and/or ILC will discover the triplet Higgs bosons.
 In particular, we concentrate on neutrino parameters which cannot be
 measured in oscillation experiments, the light neutrino mass ordering, the mass of the lightest
 neutrino and the Majorana phases $\alpha_1$ and $\alpha_2$. 
 Neutrino mass hierarchy patterns at colliders have been previously studied in \cite{Chun:2003ej}.
 First we derive analytical 
 expressions for those quantities which are functions of the doubly charged Higgs 
 branching fractions to different flavour combinations of charged lepton pairs, 
 $\Phi^{\pm\pm}\to \ell_i \ell_j,$ $i,j=e, \mu, \tau.$ 
 Thus neutrino physics at colliders turns out to be just a 
 counting experiment  of lepton flavours.
 This simplifies the life in particular at LHC experiments which, in general,  have 
 larger measurement errors than at ILC.
  The analytical results are first derived assuming the
 tri-bi-maximal mixing for neutrinos, which predicts $\sin \theta_{13}=0,$ and extended later to 
 non-zero values of $\sin \theta_{13}.$ After that we study to what precision those quantities can be measured in realistic  experiments. Finally we demonstrate that combining positive 
 colliders signals of this scenario with the possible measurement of the neutrino
 mass matrix entry $(m_\nu)_{ee}$ would allow one to determine separately the size of 
 triplet Yukawa couplings and the $B-L$ breaking VEV of the triplet. Thus one can 
 entirely probe the neutrino mass generating mechanism at terrestrial experiments.
 
 We find that there are distinctive flavour signals which indicate certain patterns of neutrino
 mass matrix. For example, very few electrons in  $\Phi^{--} $ decays definitely
 points towards normally hierarchical light neutrinos. There is theoretical ambiguity in
 determination of the Majorana phases and only a combinations of them can be measured.
 However, in the case of very hierarchical neutrino mass spectrum one of the Majorana phases
 is effectively decoupled from physics and one can, in principal, measure the magnitude
 of the physical phase. Although the experimental errors in determining those quantities
 may turn out to be quite large in general, we show that there exist scenarios which
 can already be fully solved at LHC. In the optimistic scenarios 
 the branching fractions of the doubly charged Higgs boson decays can be used to $(i)$ determine the neutrino mass hierarchy; $(ii)$ estimate the mass of the lowest neutrino state; $(iii)$ estimate the Majorana phases of CP violation; $(iiii)$ measure the value of Higgs triplet VEV.  
 We note that those measurements are also sensitive to all other neutrino parameters including the mixing angles and CP violating phase $\delta$. 
 We show that the latter two are, in principle, measurable at collider experiments.
 However, those quantities
 can be determined with much higher precision in other experiments and 
 we do not study their effects in detail in this paper.

The paper is organized as follows. 
In Section \ref{Sec2} we present the collider phenomenology of doubly charged
Higgs boson and relate the collider observables to the neutrino parameters. 
In Section \ref{Sec3} we present details of the analysis of neutrino parameter measurements 
at colliders. In Section \ref{Sec4} we discuss the possibility of measuring the triplet Higgs VEV and
determination of the full neutrino mass matrix. Finally we conclude in Section \ref{Sec5}.

\section{Phenomenological setup} \label{Sec2}

In this work we assume that the SM particle spectrum is extended by  a
scalar multiplet $\Phi$ with the $SU(2)_L\times U(1)_Y $ quantum numbers 
$\Phi\sim (3,2)$. We also assume that its mass is below ${\cal O}(1)$~TeV
and the pair production processes at colliders,
\bea
pp\to\Phi^{++}\Phi^{--} \quad \mrm{and} \quad e^+e^-\to\Phi^{++}\Phi^{--},
\label{production}
\eea 
are kinematically allowed. Such a scenario is realized, for example, in
the little Higgs models \cite{Arkani-Hamed:2001ca,Cheng:2001vd,Arkani-Hamed:2001nc,Arkani-Hamed:2002qy}. 

The triplet couples to leptons via the Lagrangian
\begin{equation}
L=i\bar \ell^c_{Li} \tau_2  Y_\Phi^{ij} (\tau\cdot \Phi) \ell_{Lj} 
+ h.c. ,
\label{L}
\end{equation}
where $(Y_\Phi)_{ij}$ are the Majorana Yukawa couplings of the triplet to the
lepton generations $i,j=e, \mu, \tau$. If  the neutral
component of  triplet acquires a VEV $v_\Phi$,
the non-zero neutrino mass matrix  is generated via
\begin{equation}
(m_\nu)_{ij} = 2(Y_\Phi)_{ij} v_\Phi.
\label{mnu}
\end{equation}
To avoid the existence of phenomenologically unacceptable Majoron   
the $B-L$ breaking VEV $v_\Phi$ cannot occur spontaneously. Instead
it should be induced effectively via the coupling of $\Phi$ to the 
SM Higgs doublet $H$ as $\mu \Phi^0 H^0 H^0$, where the dimensionful 
parameter $\mu$ breaks $B-L$ explicitly~\cite{Ma:1998dx}. 
Because in the limit $\mu\to 0$
the symmetry of the model is enhanced, it is natural that $\mu$
is a small parameter. Model building in this direction ~\cite{Ma:2000wp,Ma:2000xh,Sahu:2007uh}
is beyond the scope of the present analyses. Indeed, the above described scenario 
 is consistent with the observation that
neutrino masses are much smaller than the masses of other SM fermions.
Thus the smallness of neutrino masses is explained by be smallness of 
$v_\Phi$ and the Yukawa couplings $(Y_\Phi)_{ij}$ can be of order SM
Yukawa couplings. The most stringent constraint on them arises from 
non-observation of the muon decay $\mu\to eee$ which implies
$Y_{ee}Y_{e\mu}^* < 2\cdot 10^{-5}$ for $m_\Phi=1$ TeV \cite{Yue:2007kv}.

It is important to emphasize that the precise values of $(Y_\Phi)_{ij}$ are not 
relevant for the collider physics we consider in this work.
The relationship between neutrino parameters and doubly charged Higgs boson decays comes from the fact that the Yukawa coupling matrix of doubly charged Higgs to leptons is proportional to the Majorana mass matrix as given by Eq.\eqref{mnu}. Thus, to establish this connection experimentally,
observable rates of the leptonic branching fractions must exist.

The decay width of doubly charged Higgs to the  corresponding leptonic channel is given by
\begin{equation}
\Gamma_{ij}\equiv \Gamma(\Phi^{\pm\pm}\rightarrow\ell_i^\pm\ell_j^\pm) =\left\{\begin{array}{l}
\frac1{8\pi}|(Y_\Phi)_{ii}|^2 m_{\Phi^{\pm\pm}}\qquad i=j, \\
\\
\frac1{4\pi}|(Y_\Phi)_{ij}|^2 m_{\Phi^{\pm\pm}}\qquad i\ne j,
\end{array}\right.
\label{dw}
\end{equation}
and the decay width to the $WW$ channel is
\begin{equation}
\Gamma_{WW}\equiv \Gamma(\Phi^{\pm\pm}\rightarrow W_L^\pm W_L^\pm)=\frac{g_L^4v_\Phi^2 m_{\Phi^{\pm\pm}}}{16\pi m_{W_L^\pm}^2}\left(\frac{3m_{W_L^\pm}^2}{m_{\Phi^{\pm\pm}}^2}+\frac{m_{\Phi^{\pm\pm}}^2}{4m_{W_L^\pm}^2}-1\right)\left(1-\frac{4m_{W_L^\pm}^2}{m_{\Phi^{\pm\pm}}^2}\right)^{1/2}
\equiv k v^2_\Phi.
\label{dwW}
\end{equation}

The branching ratio of $\Phi^{\pm\pm}$ to a single leptonic channel can be calculated using the decay widths
\begin{equation}
\mrm{\mrm{BR}}_{ij} \equiv \mrm{\mrm{BR}}(\Phi^{\pm\pm}\rightarrow\ell_i^\pm\ell_j^\pm)=  \frac{\Gamma_{ij}}{\Gamma_{tot}}, \label{branch}
\end{equation}
where $\Gamma_{tot} = \sum_{i\ge j} \Gamma_{ij}+\Gamma_{WW}$ is the total decay width.
Since $\Gamma_{ij}$ is directly related to neutrino mass matrix, we can derive a relation  between the $\Phi^{++}\Phi^{--}$ branching ratios that can be measured in collider experiments and the neutrino mass matrix, that contains all currently unknown neutrino parameters. The branching ratio to a single decay channel can be found combining Eqs.~\eqref{mnu}, \eqref{dw} and \eqref{branch}
\begin{equation}
\mrm{\mrm{BR}}(\Phi^{\pm\pm}\rightarrow\ell_i^\pm\ell_j^\pm)=\frac{|(m_\nu)_{ij}|^2}{\sum_{i\ge j} |(m_\nu)_{ij}|^2+4k v_\Phi^4}\label{hgss},
\end{equation}
where $(m_\nu)_{ij}$ is the neutrino mass matrix in flavor basis.  Eq.~\eqref{hgss} shows the direct relationship between neutrino parameters and the  $\Phi^{\pm\pm}$ branching ratios. 

The neutrino mass matrix can be diagonalized by unitary leptonic mixing matrix $U$,
\begin{equation}
m_\nu=U^* m_\nu^D U^\dagger,
\end{equation}
where the diagonalized neutrino mass matrix $m_\nu^D$ is given by
\[
m_\nu^D=\left(
\begin{array}{ccc}
m_1 & 0 & 0 \\
0 & m_2 & 0 \\
0 & 0 & m_3
\end{array}
\right).
\]
Here $m_1$, $m_2$ and $m_3$ represent the masses of neutrino mass eigenstates $\nu_1$, $\nu_2$ and $\nu_3$. $\nu_1$ and $\nu_2$ masses differ by $\Delta m^2_{sol}= 7.92(1\pm0.09)\times 10^{-5} \mrm{eV}^2$ measured by solar oscillation experiments \cite{Fogli:2006yq} and $m_1<m_2$. The third eigenstate $\nu_3$ is separated from the first two by splitting $\Delta m^2_{atm} = 2.6(1^{+0.14}_{-0.15})\times 10^{-3} \mrm{eV}^2$  \cite{Fogli:2006qg} and can be heavier or lighter than the solar pair. The two possibilities are called normal and inverted spectrum, respectively. The third possibility -- nearly degenerate masses -- appears when the lowest neutrino mass is large compared to the measured mass differences and $m_1\approx m_2\approx m_3$. Cosmology implies that neutrinos are lighter than about $0.2$ eV \cite{Lesgourgues:2006nd}.

Since  we have assumed that there are only three Majorana neutrinos, $U$ is a $3\times3$  mixing matrix that depends on three mixing angles and three phases and can be parameterized   as
\begin{equation}
U=\left(\begin{array}{ccc}
1 & 0 & 0 \\
0 & c_{23} & s_{23}\\
0 & -s_{23} & c_{23}
\end{array}\right)
\left(\begin{array}{ccc}
c_{13} & 0 & s_{13}e^{-i\delta} \\
0 & 1& 0\\
-s_{13}e^{i\delta} & 0 & c_{13}
\end{array}\right)
\left(\begin{array}{ccc}
c_{12} & s_{12} & 0 \\
-s_{12} & c_{12}& 0\\
0 & 0 & 1
\end{array}\right)
\left(\begin{array}{ccc}
e^{i\alpha_1} & 0 & 0 \\
0 & e^{i\alpha_2}& 0\\
0 & 0 & 1
\end{array}\right),\label{U}
\end{equation}
where $c_{ij} \equiv \cos{\theta_{ij}}$, $s_{ij}\equiv\sin{\theta_{ij}}$ and $\theta_{ij}$ denote
 the mixing angles. The quantities $\delta$, $\alpha_1$ and $\alpha_2$ are CP violating phases. 
 $\delta$ is the Dirac phase and characterizes CP violation regardless of the character of neutrinos. $\alpha_1$ and $\alpha_2$ are called Majorana phases and are physical only if neutrinos are Majorana particles. If the neutrinos were Dirac fermions, both Majorana phases could be absorbed by appropriately redefining the neutrino fields, and the only observable CP violation parameter would be the Dirac phase $\delta$. Also note that $\delta$ appears in the mixing matrix only as $\sin{\theta_{13}} e^{i\delta}$ - so the influence of $\delta$ crucially depends on the value of $\theta_{13}$ and has physical consequences only if $\theta_{13}$ is non-zero. 

The mixing matrix contains six independent parameters: three mixing angles ($\theta_{13}$, $\theta_{23}$ and $\theta_{12}$) and three phases ($\alpha_1$, $\alpha_2$ and $\delta$). 
Mixing angles are known from the global fit to neutrino oscillation data and are given by (2$\sigma$ errors) \cite{Fogli:2006yq,Fogli:2006qg}
\begin{equation}
\sin^2{\theta_{12}}=0.314(1_{-0.15}^{+0.18})\text{  ,         }\sin^2{\theta_{23}}=0.45(1^{+0.35}_{-0.20})\text{  ,        }\sin^2{\theta_{13}}=0.8^{+2.3}_{-0.8}\times10^{-2}. \label{nmix}
\end{equation}
Up to now no experiment has been able to determine the values of phases so that
 \begin{equation}
 \delta\text{,  } \alpha_1\text{,   }\alpha_2\text{   } \in[0,2\pi],
\end{equation}
remain unconstrained.

So far we have shown that doubly charged Higgs boson decay statistics is directly related to neutrino parameters as given by Eq.~\eqref{hgss} and depends on neutrino mass matrix $m_\nu^D$ and mixing matrix $U$. Additional information can be acquired from the neutrinoless double beta decay  experiments, that independently probe the absolute value of the $(m_\nu)_{ee}$ entry 
in neutrino Majorana mass matrix.
Such relations allow direct measurements of neutrino parameters in particle collider experiments.

\section{Measuring neutrino parameters at colliders} \label{Sec3}

Doubly charged Higgs boson has 6 different leptonic decay channels. Branching ratios to these channels are functions of neutrino parameters according to Eq.~\eqref{hgss}. We have fixed the values of mass differences $\Delta m^2_{atm}$ and $\Delta m^2_{sol}$ in the subsequent calculations, as they have been measured with a good precision in neutrino oscillation experiments. With such an assumption
 we can write an equation system of six independent equations that relates branching ratios of six different $\Phi^{\pm\pm}$ leptonic decay channels with unknown neutrino parameters,
\begin{equation}
\mrm{\mrm{BR}}_{ij}=f_k(m_0, \mathrm{sign}(\Delta m_{atm}), \theta_{13},\theta_{23}, \theta_{12}, \delta,\alpha_1, \alpha_2),\label{np}
\end{equation}
where $m_0$ represents the mass of the lowest neutrino mass eigenstate ($m_1$ or $m_3$ for normal or inverted mass spectrum, respectively), $k=1, ... , 6$ and $i,j=e,\mu,\tau$. 
In the subsequent analyses we have used relations between the leptonic branching ratios instead of their absolute values. This method 
 is independent of the possible $\Phi^{\pm\pm}$ decay to $WW$ channel, which is more complicated to measure accurately at LHC. In such an approach we simply count the events of $\Phi^{\pm\pm}$ decays to different channels and calculate their relative differences. As a result we have five independent equations. In order to solve this equation system with respect to unknown neutrino parameters, we have to fix at least some of them. Consequently we can solve the equation system \eqref{np} for different neutrino parameters and obtain them as functions of  the $\Phi^{\pm\pm}$ leptonic branching ratios $\mrm{\mrm{BR}}_{ij}$.

\subsection{Results for the tri-bi-maximal mixing}\label{3.1}

Since approximate values of neutrino mixing angles are known from oscillation experiments and the precision of measurements is expected to be increased in upcoming years \cite{fixtheta}, we fix their values in most of our analyses. 
 We have chosen to follow the tri-bi-maximal model \cite{Harrison:1999cf} .  It has been proposed  that the combined existing data from neutrino oscillations point to a specific form of the lepton mixing matrix with effective bi-maximal mixing of $\nu_\mu$ and $\nu_\tau$ at the atmospheric scale and effective tri-maximal mixing at the solar scale - hence denoted as tri-bi-maximal mixing.  The tri-bi-maximal mixing predicts 
 \begin{equation}
\sin^2{\theta_{12}}=\frac{1}{3}\text{ ,    } \sin^2{\theta_{23}}=\frac{1}{2}\text{ ,    } \sin^2{\theta_{13}}=0,\label{tribi}
 \end{equation}
which are perfectly compatible with the present experimental uncertainties given by Eq.~\eqref{nmix}. The main aim of this paper is to provide information about the Majorana phases and absolute values of neutrino masses. In the tri-bi-maximal model the CP violating phase $\delta$ is not physical due to the zero value of $\theta_{13}$ and the only remaining unknown variables are the lowest neutrino mass $m_0$, neutrino hierarchy i.e. $\mathrm{sign}(\Delta m_{atm})$ and Majorana phases $\alpha_1$ and $\alpha_2$.

Having fixed the mixing angles according to Eq.~\eqref{tribi}, we end up with four independent equations for branching ratios, since $\mrm{\mrm{BR}}_{e\mu}=\mrm{\mrm{BR}}_{e\tau}$ and $\mrm{\mrm{BR}}_{\mu\mu}=\mrm{\mrm{BR}}_{\tau\tau}$. If a measurement would show that these branching ratios are not equal, this is a clear indication that the tri-bi-maximal model has to be modified.
As we are using the relations between branching ratios for the calculations, the number of independent equations is reduced to three. Such equation system can be solved with respect to three unknown parameters: the lowest neutrino mass $m_0$ and Majorana phases $\alpha_1$ and $\alpha_2$.
 We show how the mass of the lowest neutrino mass eigenstate, neutrino mass hierarchy and the difference of two Majorana phases $|\Delta\alpha|$ can be uniquely determined from the relation \eqref{hgss}.  Unique solutions for $\alpha_1$ and $\alpha_2$ are not determined by the 
$\Phi^{++}\Phi^{--}$ branching ratios, and two sets of degenerate solutions are found.

\subsubsection{Neutrino hierarchy and the lowest neutrino mass}

First  we consider the equation system given by Eq.~\eqref{np} with the fixed tri-bi-maximal mixing angles. 
 For the neutrino mass hierarchy and lowest neutrino mass determination we combine the branching ratios of $\mu\mu$, $\mu\tau$, $ee$ and $e\mu$ channels.  After some simple algebra we  find a relation between these branching ratios that depends only on neutrino masses and is independent of the Majorana phases,
\begin{equation}\label{m1y}
C_1 \equiv 
\frac{2\mrm{BR}_{\mu\mu}+\mrm{BR}_{\mu\tau}-\mrm{BR}_{ee}}{\mrm{BR}_{ee}+\mrm{BR}_{e\mu}}=
\frac{-m_1^2+m_2^2+3m_3^2}{2m_1^2 + m_2^2}.
\end{equation}
Here and onwards in this paper $C_x$ denote  constant dimensionless parameters which 
can be measured in experiments.

The mass hierarchy can be easily determined by simply measuring the value of $C_1$ that is independent of the values of $\alpha_1$ and $\alpha_2$. It can be found that $C_1$ uniquely determines the mass hierarchy as follows:
\begin{itemize}
\item $C_1>1$ -- normal mass hierarchy,
\item $C_1<1$ -- inverted mass hierarchy,
\item $C_1\approx 1$ -- degenerate masses.
\end{itemize}
After the mass hierarchy measurement we can solve Eq.~\eqref{m1y} for either normal or inverted mass hierarchy.  
For the normal mass hierarchy $m_1$ is the lowest mass state, $m_2^2=m_1^2+\Delta m^2_{sol}$ and $m_3^2=m_1^2+\Delta m^2_{sol}+\Delta m^2_{atm}$. After substituting $m_2$ and $m_3$, we get the following equation that can be solved with respect to $m_1$,
\begin{equation}\label{norm}
m_1^2 = \frac{\Delta m^2_{sol}(4 - C_1)+ 3\Delta m_{atm}^2}{3( C_1 - 1)}.
\end{equation}

Alternatively, for the inverted mass hierarchy $m_3$ is the lowest mass state, $m_2^2=m_3^2+\Delta m^2_{atm}$ and $m_1^2 = m_3^2+\Delta m^2_{atm}-\Delta m^2_{sol}$. After the substitutions, Eq.~\eqref{m1y} can be solved with respect to $m_3$ as follows:
\begin{equation}
m_3^2 = \frac{\Delta m_{sol}^2(1+2 C_1) - 3C_1 \Delta m_{atm}^2}{3(C_1-1)}.
\label{invmass}
\end{equation}

\begin{figure}[t]
\begin{center}
\includegraphics[width=0.49\textwidth]{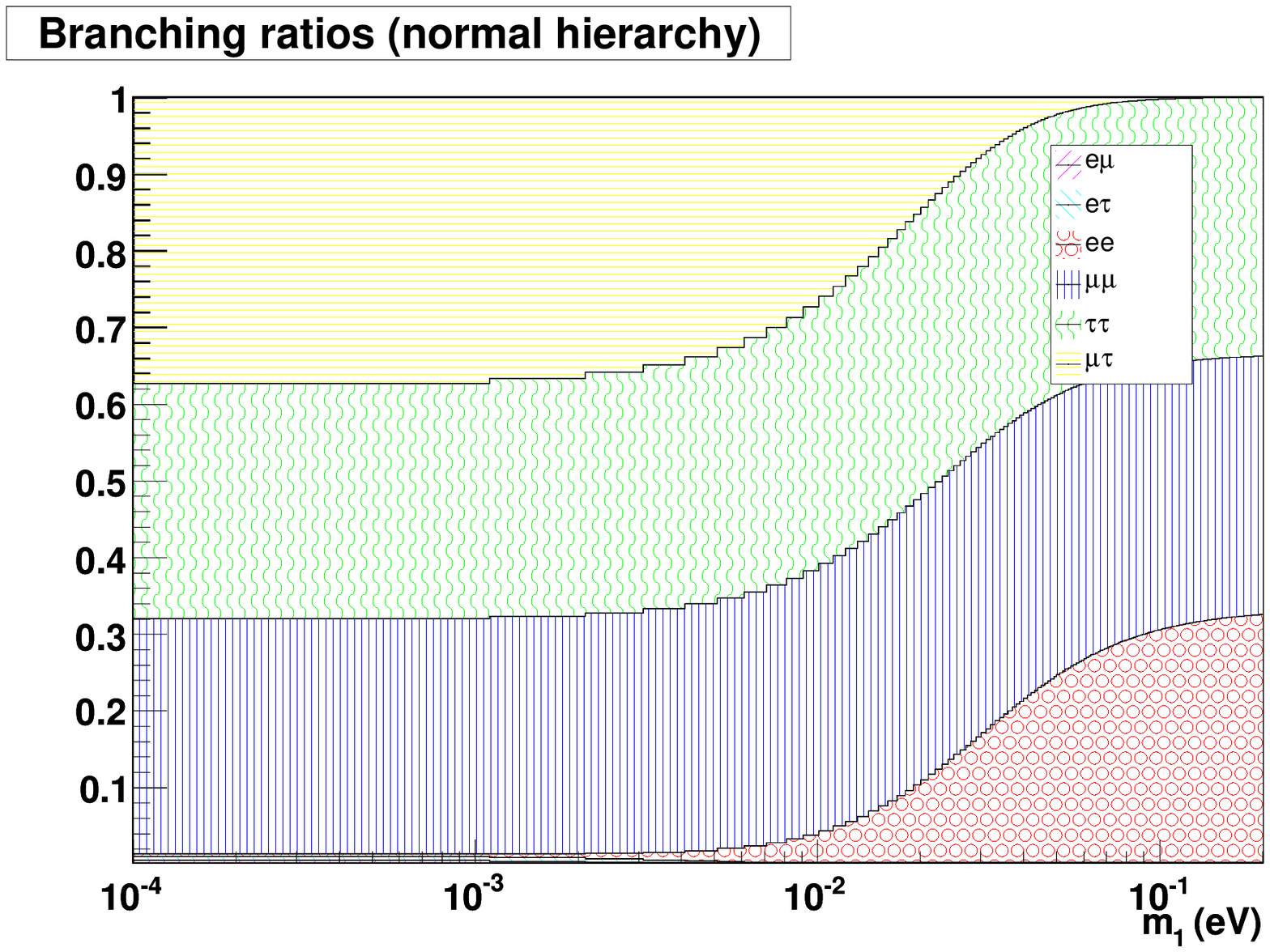}
\hfill
\includegraphics[width=0.49\textwidth]{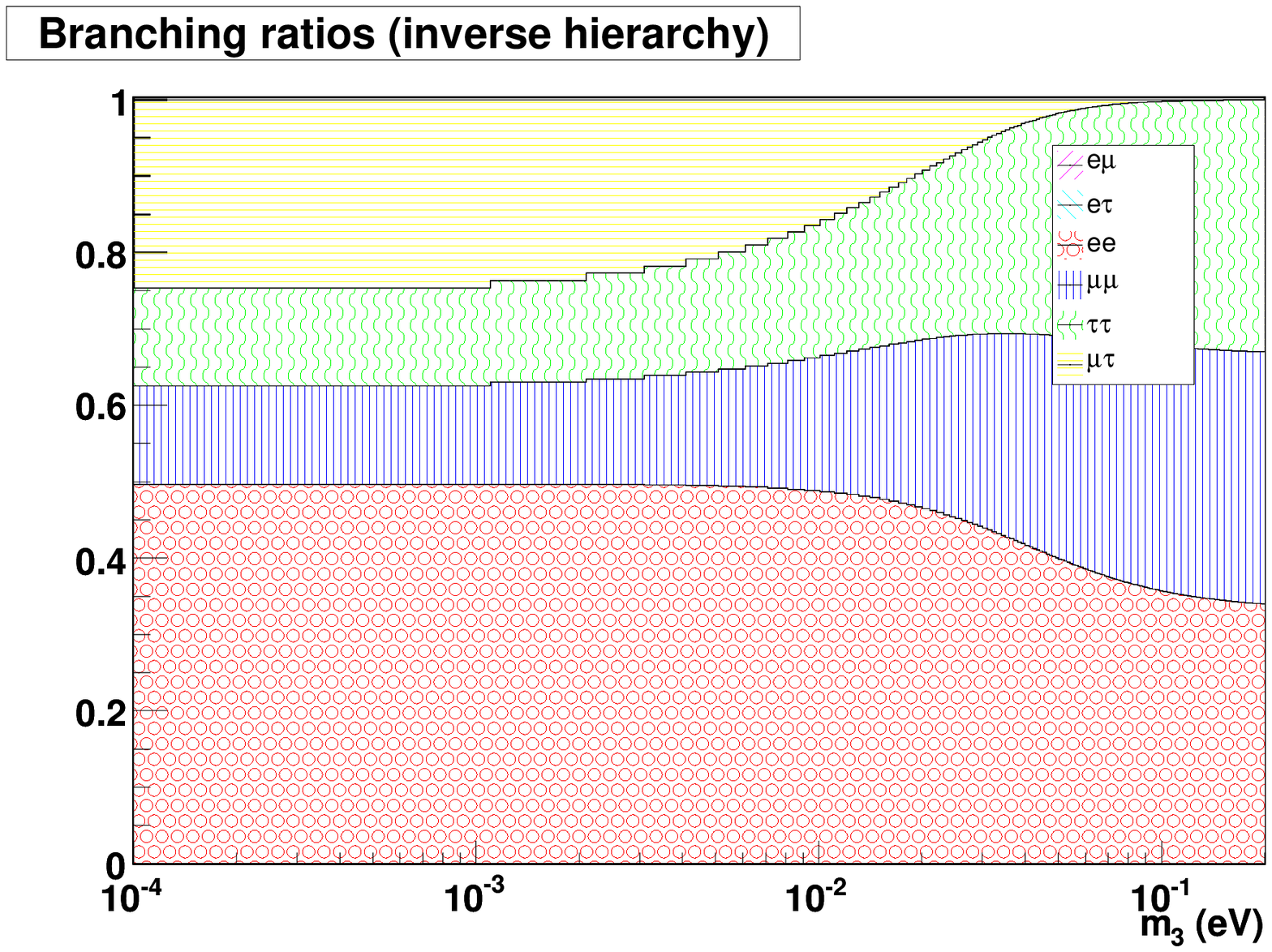}
\caption{Distribution of the $\Phi^{++}$ leptonic branching ratios as a function of the lightest neutrino mass. The left (right) panel corresponds to the normal (inverted) mass hierarchy. For nearly degenerate masses the two possibilities imply almost the same result.}
\label{mass}
\end{center}
\end{figure}

For nearly degenerate masses ($m_1>0.1 \mrm{eV}$) accurate measurement of the lowest neutrino mass requires very good experimental precision (which is not likely to be achieved at LHC) because the branching ratios are increasingly less mass dependent for larger mass values. 
This is demonstrated in Figure \ref{mass} which presents the dependency of doubly charged Higgs 
 branching ratios on the lightest neutrino mass for the normal and inverted mass hierarchies. We have assumed a real mixing matrix i.e. fixed Majorana phases to zero. The $e\mu$ and $e\tau$ channels have only vanishingly small contributions for $\alpha_1=\alpha_2 = 0$, but are increased for non-zero values of the Majorana phases. The branching ratio to $ee$ channel is a especially good characteristic for mass hierarchy determination that varies greatly depending on the hierarchy and the neutrino mass. This branching ratio is negligible for the normal mass hierarchy with very small mass while it is the dominant decay channel for the inverted mass hierarchy. If the mass of the lightest state increases, both the normal and inverted hierarchies have almost the same distribution of branching ratios, $\Phi^{\pm\pm}$ decay to $ee$, $\mu\mu$ and $\tau\tau$ with nearly equal probabilities while the decays to other channels are negligible. This indicates the degenerate masses.

\subsubsection{Majorana phases}
If the neutrino masses are measured as shown in the previous section, we can determine the values of Majorana phases in a similar way. Once again we use the tri-bi-maximal values for all mixing angles and combine  expressions from the equation system \eqref{np}. We first determine the difference between the Majorana phases $\Delta\alpha = |\alpha_1 - \alpha_2|$.  Using a relation between the $ee$ and $e\mu$ decays channels we obtain 
\begin{equation}
C_2\equiv 
\frac{\mrm{BR}_{e\mu}}{\mrm{BR}_{ee}}=
\frac{2(m_1^2+ m_2 ^2-2m_1m_2\cos{\Delta\alpha})}{4 m_1^2+ m_2^2+ 4m_1m_2\cos{\Delta\alpha}}.
\end{equation}
From this expression we can find separate solutions for the different mass hierarchies. For the normal hierarchy $\Delta\alpha$ can be found to be
\begin{equation}\label{normalpha}
\Delta\alpha =\arccos{\left(\frac{ (4 - 5 C_2)m_1^2 + (2 - C_2)\Delta m^2_{sol}}{4(1+C_2) m_1\sqrt{m_1^2+\Delta m^2_{sol}}}\right)},
\end{equation}
while  for the inverted hierarchy we find
\begin{equation}
\Delta\alpha = \arccos{\left(\frac{2(2C_2 -1)\Delta m^2_{sol} + (4-5C_2)(\Delta m^2_{atm}+m_3^2)}{4(1+C_2)\sqrt{(\Delta m^2_{atm}+m_3^2)(m_3^2+\Delta m_{atm}^2 - \Delta m_{sol}^2)}}\right)},
\end{equation}
which can be approximated as
\begin{equation}\label{invalpha}
\Delta\alpha =\arccos{\left( \frac{4-5C_2}{4(1+C_2)}\right)}+{\cal O}\left(\frac{\Delta m_{sol}^2}{\Delta m_{atm}^2}\right).
\end{equation}
For the inverted hierarchy, up to small corrections, the equation for $\Delta\alpha$ is independent of the value of $m_3$. This means that Eq. \eqref{invalpha} is valid both for the inverted hierarchy and degenerate mass spectrum. The solution for the normal hierarchy given by Eq. \eqref{normalpha} contains the lowest neutrino mass which must be measured previously with an acceptable precision.

We found that  $|\Delta\alpha|$ can be uniquely determined up to a sign uncertainty  $\mathrm{sgn}(\alpha_1 - \alpha_2)$ since cosine is an even function. In order to find a solution that separately determines $\alpha_1$ and $\alpha_2,$  we use the expression for $\Delta\alpha$ given either by Eq.~\eqref{normalpha} or Eq.~\eqref{invalpha} together with the definition of cosine of the difference of angles and construct the equation system of two independent equations,
 \bea
C_3 & \equiv & 
\frac{2{\mrm{BR}_{\mu\mu}}-{\mrm{BR}_{\mu\tau}}}{\mrm{BR}_{ee}+\mrm{BR}_{e\mu}}  = 
\frac{2m_3(\cos{\alpha_1} m_1+2\cos{\alpha_2}m_2)}{2m_1^2+m_2^2} ,
\nonumber
\\
\cos{\Delta\alpha}&= & \cos{\alpha_1}\cos{\alpha_2}+\sin{\alpha_1}\sin{\alpha_2} .
\eea 
Unfortunately such an equation system does not have a unique solution due to the uncertainty in $\mathrm{sgn}(\alpha_1 - \alpha_2)$ and two sets of degenerate solutions for $\alpha_1$ and $\alpha_2$ are found, one of which is correct for $\alpha_1>\alpha_2$ and the other corresponds to $\alpha_2>\alpha_1$. It is not possible to tell only from the collider data which angle is bigger and which of the solutions is correct. In the following we present the effect of the Majorana phases to branching ratios for three different mass hierarchies and discuss the consequences of non-vanishing $\theta_{13}.$

\subsection{Measuring Majorana phases for different mass hierarchies}\label{appr}

 In this section we study some well motivated particular cases of neutrino mass parameters
 which can be well measured at LHC. Those are:
\begin{itemize}
\item Normal mass hierarchy, $m_1 = 0$. Branching ratios are independent of $\alpha_1$; $\alpha_2$ can be determined.
\item Inverted mass hierarchy, $m_3 = 0$. Branching ratios are independent of absolute values of Majorana phases; $\Delta\alpha = |\alpha_1 -\alpha_2|$ can be determined.
\item Nearly degenerate masses, $m_1\approx m_2\approx m_3 = m$. Expressions for branching ratios become independent of $m$.
\end{itemize}

We have kept the mixing angles fixed to tri-bi-maximal values unless stated otherwise.

\subsubsection{Normal hierarchy, $m_1=0$}

In this case the doubly charged Higgs branching ratios are independent of $\alpha_1$ and we can determine $\alpha_2$. Branching ratios to the decay channels that involve electrons can be neglected and,  for expressing the solutions, we use the branching ratios to $\mu\mu$ and $\mu\tau$ channels. Relation between
 these channels gives the following equation with $\alpha_2$ as the only unknown parameter,
\begin{equation}
C_4\equiv  \frac{\mrm{BR}_{\mu\mu}}{\mrm{BR}_{\mu\tau}}=
\frac{13\Delta m^2_{sol}+9\Delta m^2_{atm}+12\cos{\alpha_2}\sqrt{\Delta m^2_{sol}}\sqrt{\Delta m^2_{sol}+\Delta m^2_{atm}}}{2(13\Delta m^2_{sol}+9\Delta m^2_{atm}-12\cos{\alpha_2}\sqrt{\Delta m^2_{sol}}\sqrt{\Delta m^2_{sol}+\Delta m^2_{atm}})}.
\end{equation}
This can be solved uniquely for $\alpha_2$ as
\begin{equation}
\alpha_2=\arccos{\left( \frac{(2C_4-1)(13\Delta m^2_{sol}+9\Delta m^2_{atm})}{12(1+2C_4)\sqrt{\Delta m^2_{sol}}\sqrt{\Delta m^2_{sol}+\Delta m^2_{atm}}}\right)}.
\end{equation}

\begin{figure}[t]
\begin{center}
\includegraphics[width=0.49\textwidth]{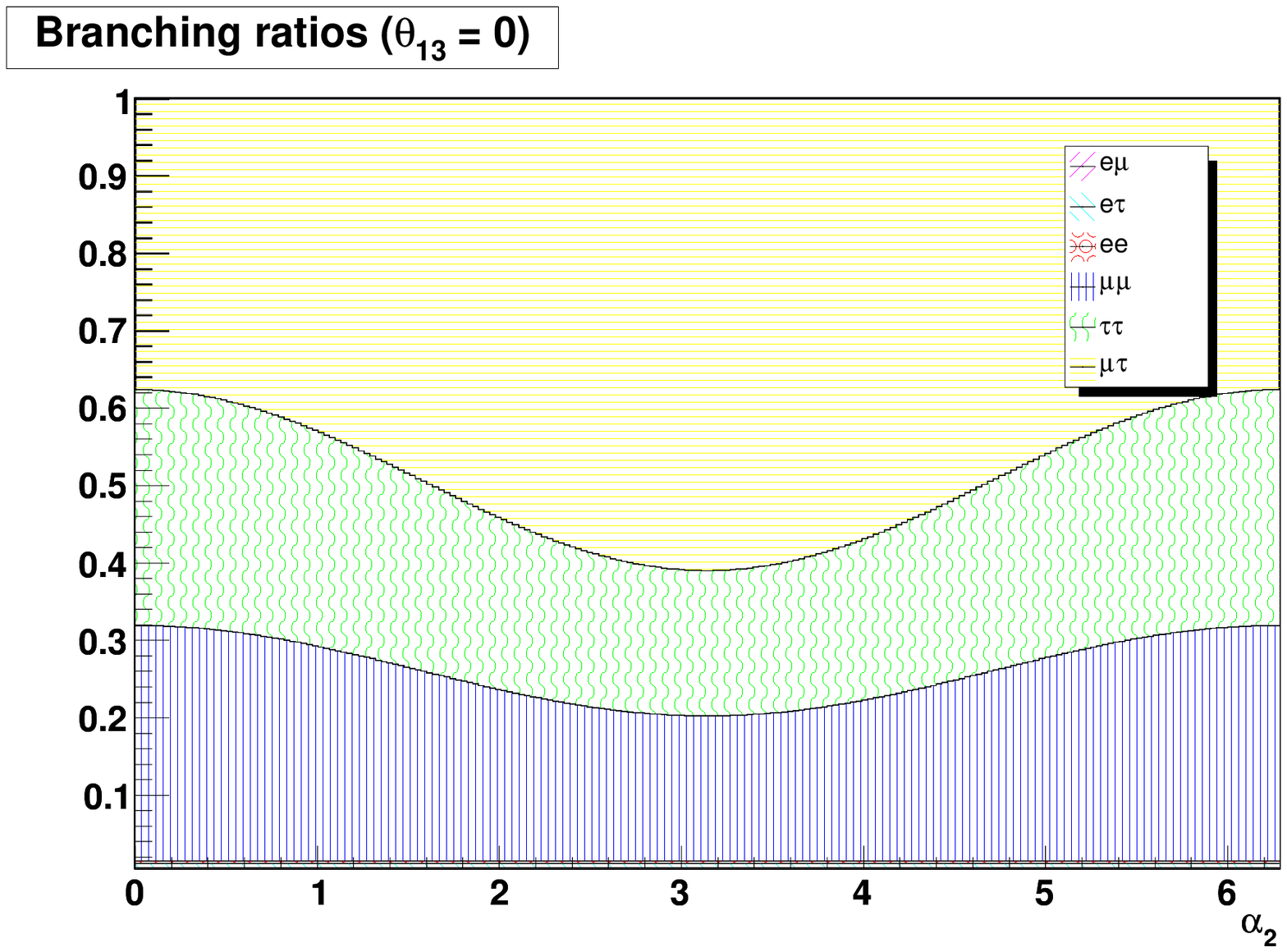}
\hfill
\includegraphics[width=0.49\textwidth]{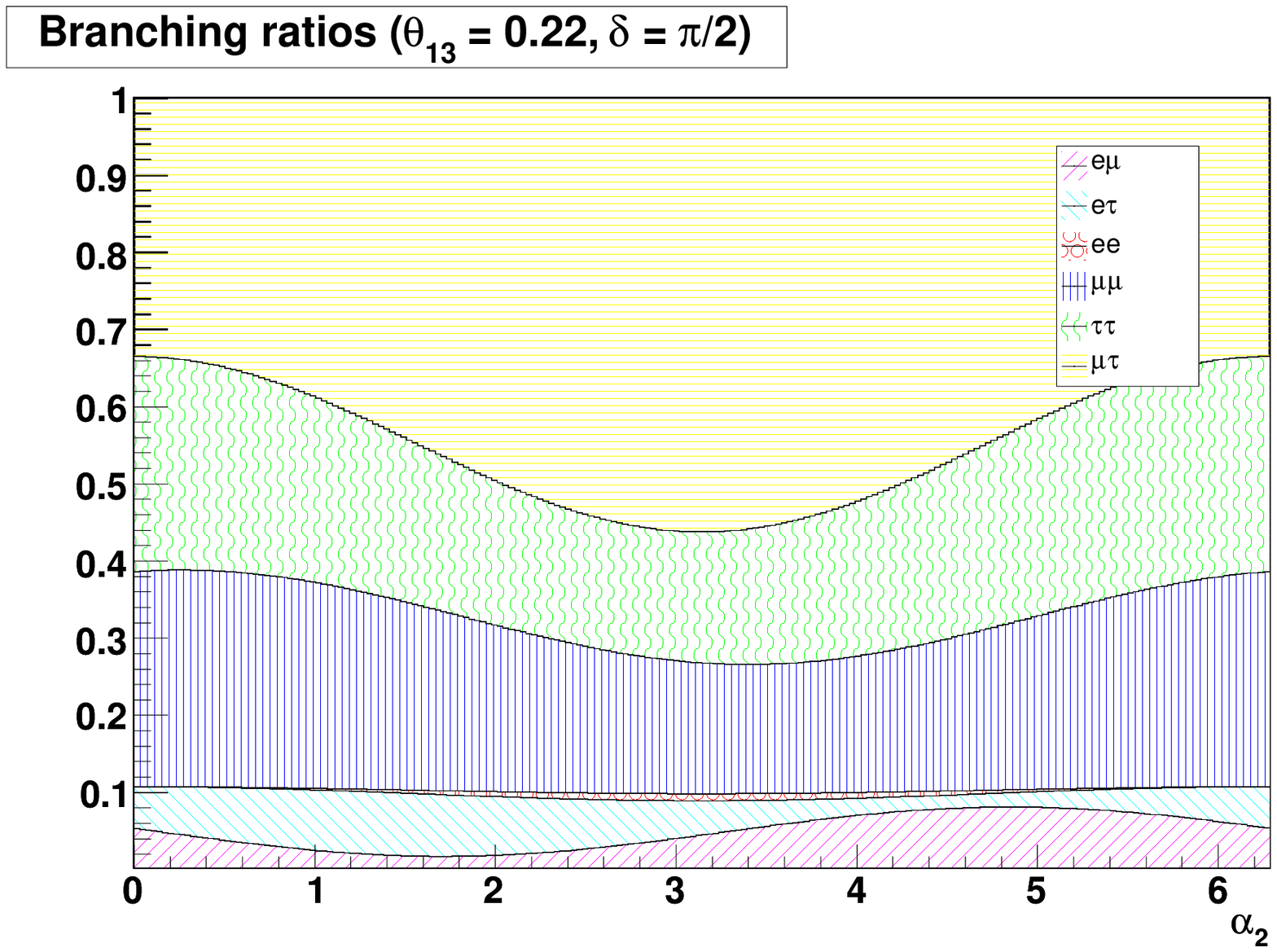}
\caption{Distributions of the branching ratios as a function of $\alpha_2$. The left panel corresponds to
 $\theta_{13}=0$ with the $ee$, $e\mu$ and $e\tau$ channels giving nearly negligible contributions. When $\theta_{13}$ is non-zero (the right panel), small branching ratios to $e\mu$ and $e\tau$ channels can be measured. Non-zero $\delta$ in the right panel causes the slight asymmetry with respect to $\alpha_2 = \pi.$}
\label{fn}
\end{center}
\end{figure}

The distribution of branching ratios for the tri-bi-maximal mixing angles is shown in the left panel of Figure \ref{fn}. The dominant decay channels are $\mu\mu$, $\tau\tau$ and $\mu\tau$. Decays including electrons can be neglected, since the branching ratios to the $ee$, $e\mu$ and $e\tau$ channels are suppressed by $\Delta m_{sol}^2/\Delta m_{atm}^2$ 
that is small compared to the relevant terms in other decay channels.
Non-zero $\alpha_2$ causes a small variation in branching ratios, the $\mu\tau$ channel is increased for while the $\tau\tau$ and $\mu\mu$ channels are reduced proportionally.  
The right panel shows the effect of non-zero $\theta_{13}$ and $\delta$ that create small non-zero contributions to the $e\mu$ and $e\tau$ channels. In this case non-zero $\theta_{13}$ could be clearly detected.  However, those can comprise only about $10\%$ of all the decays and require high
statistics to be adequately measured at colliders. We also emphasize that the asymmetry of distributions
in this case is CP-violation effect due to non-vanishing Dirac phase $\delta.$

In conclusion, if we have identified the normal mass hierarchy with nearly zero value of $m_1$, which is being described by $\mu\mu$, $\mu\tau$ and $\tau\tau$ as the dominant decay channels, we can measure $\alpha_2$ from the ratio between $\mu\mu$ and $\mu\tau$ channels. We note that the changes in branching ratios are symmetrical with respect to $\alpha_2 = \pi$ and we always have two possible solutions. The non-zero $\theta_{13}$ and $\delta$ can create a slight asymmetry in the solutions 
due to the CP-violation and thus provide a possibility of unique determination of  $\alpha_2.$ However, this is a very small effect that requires a precision measurement and most likely can not be detected at the LHC.

\subsubsection{Inverted hierarchy, $m_3=0$}
\begin{figure}[h]
\begin{center}
\includegraphics[width=0.49\textwidth]{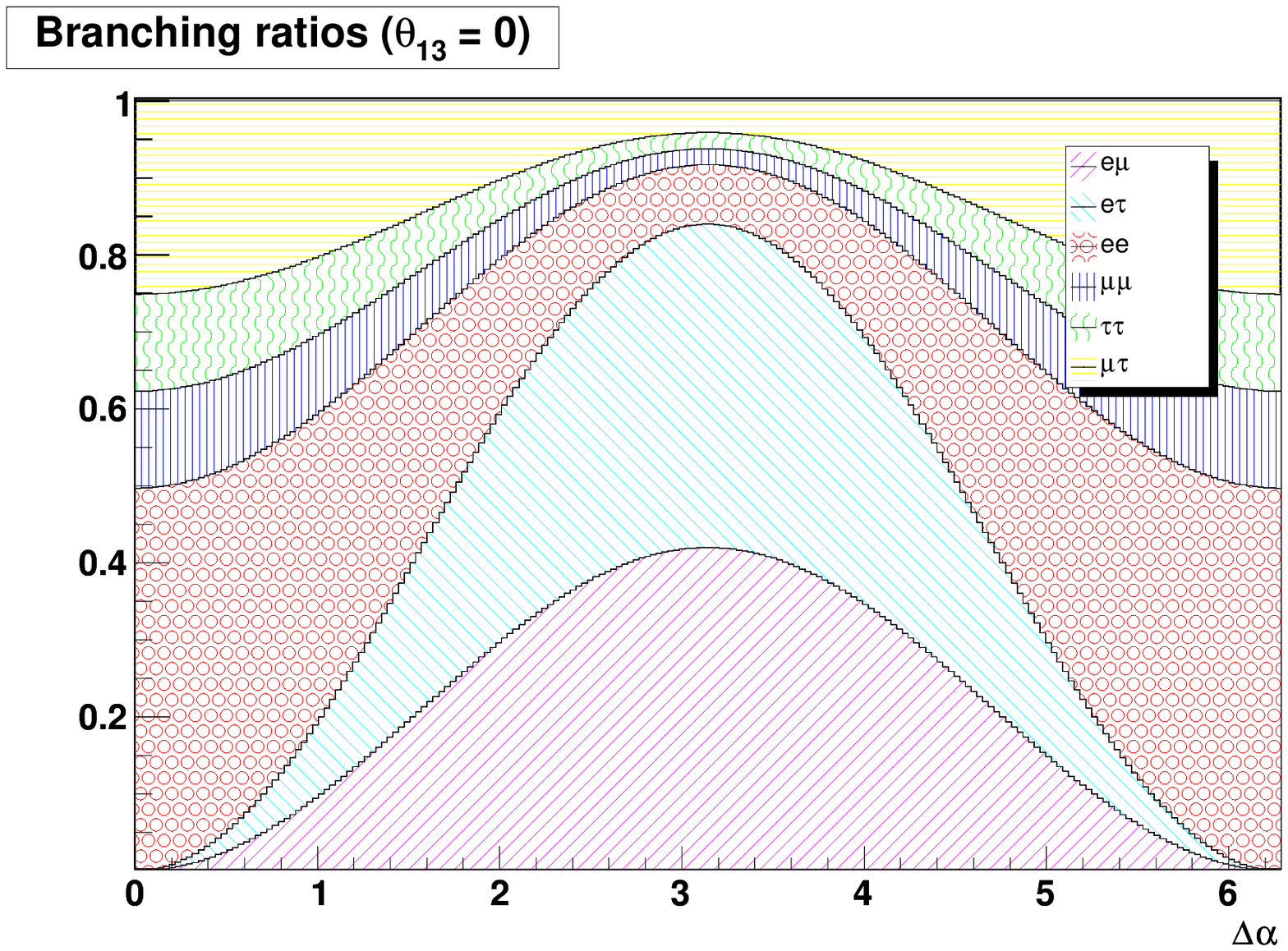}
\hfill
\includegraphics[width=0.49\textwidth]{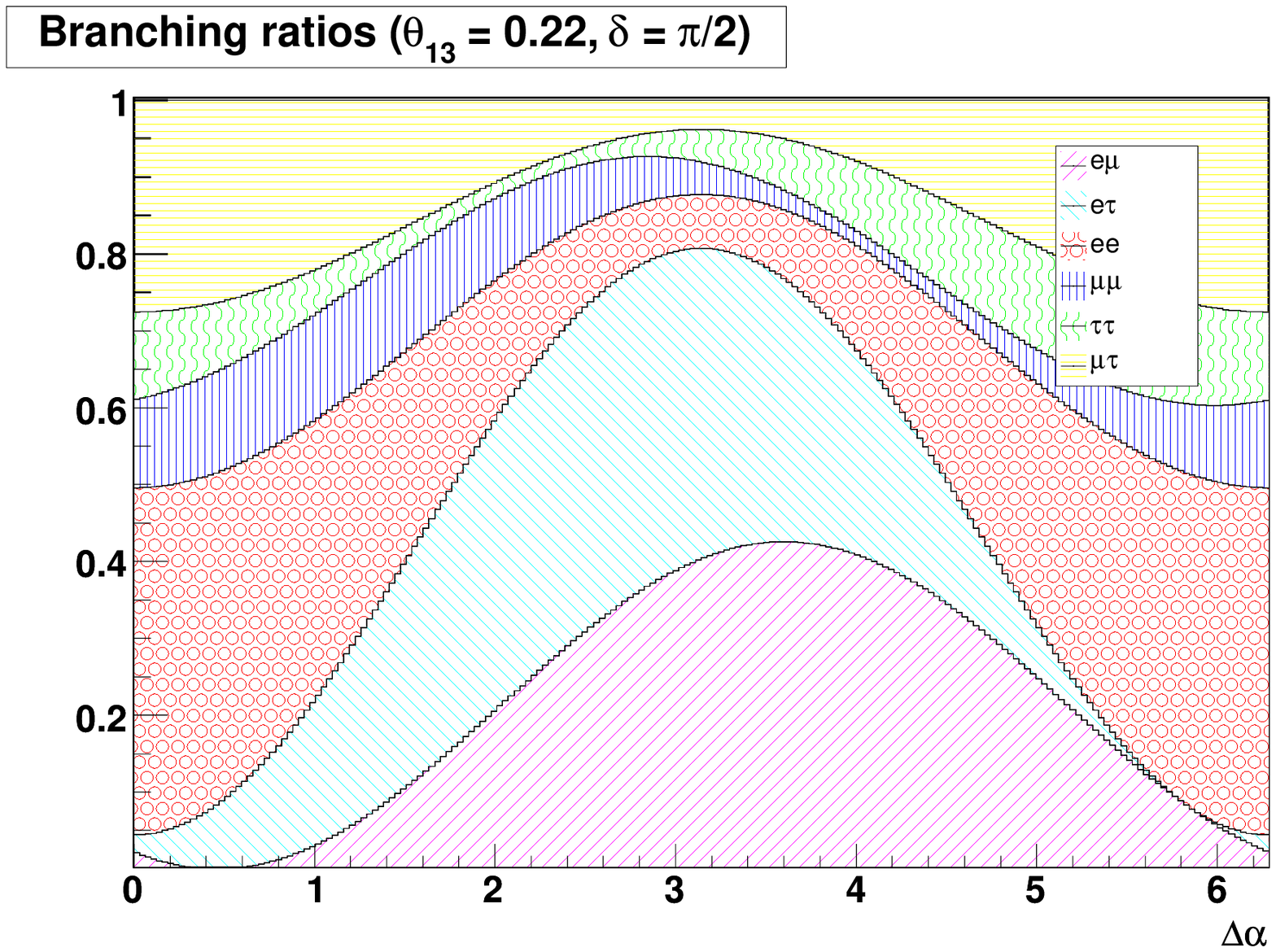}
\caption{Distribution of the $\Phi^{\pm\pm}$ branching ratios as a function of $\Delta\alpha$
for $\theta_{13}=0$ (left panel), and $\theta_{13}=0.22$, $\delta=\pi/2$ (right panel). The asymmetry
of the latter plot signals non-vanishing CP-violation.}
\label{inv}
\end{center}
\end{figure}

In this case the  branching ratios do not depend on absolute values of Majorana phases and only their relative difference $\Delta \alpha$ can be measured. We still can use  Eq.~\eqref{invalpha} to determine the value of $\Delta\alpha$. The distribution of all $\Phi^{\pm\pm}$ branching ratios as functions of $\Delta\alpha$ is presented in Figure \ref{inv}. One can see that the changes in branching ratios caused by non-zero $\Delta\alpha$ are much more prominent than the changes caused by $\alpha_2$ for normal hierarchy. When $\Delta\alpha=0$, $ee$ is the dominant decay channel and the $e\mu$ and $e\tau$ channels can have only very small contributions resulting from small non-zero $\theta_{13}$. If $\theta_{13} = 0$ is assumed, any non-zero contribution to the $e\mu$ or $e\tau$ channels would indicate non-zero value of $\Delta\alpha$. Non-zero  $\Delta\alpha$ suppresses the $ee$ channel considerably and the branching ratios to $e\mu$ and $e\tau$ channels can occupy more than $80\%$ of all leptonic decays. Branching ratio to the $ee$ channel remains non-zero in this case. The effect of non-zero  $\theta_{13}$ and CP violation angle $\delta$ is presented in the right panel of  Figure \ref{inv}. 
It makes the distributions slightly asymmetric with respect to  $\Delta\alpha = \pi$ which, in principle,
can be measured.  As in the previous case, the asymmetry is a signal of CP-violation.

\subsubsection{Degenerate masses}
When neutrino masses are large compared to the mass differences $\Delta m_{sol}$ and $\Delta m_{atm}$, all three mass states are approximately equal and we can use the model of nearly degenerate neutrino masses: $m_1\approx m_2\approx m_3 = m$. As shown before, the exact value of neutrino mass can not be determined by $\Phi^{\pm\pm}$ decay statistics, since $m$ becomes independent of branching ratios in the degenerate limit. This means, that $m$ is canceled from the expressions for branching ratios and the calculation  of the  Majorana phases is significantly simplified. 
To obtain general predictions we first assume a small value of $\theta_{13}$, so that higher order terms of the expansion can be neglected, and do not fix other parameters. 

First we can check whether any of the Majorana phases has a non-zero value.  When $\alpha_1=\alpha_2=0$, we would observe nearly equal amount of decays to $ee$, $\mu\mu$ and $\tau\tau$ channels, while all other decay channels would be suppressed:
  \begin{eqnarray}
  \mrm{BR}_{ee}&\approx&\mrm{BR}_{\mu\mu}=\mrm{BR}_{\tau\tau}=\frac{1}{3},\\
\mrm{BR}_{e\mu}&=&\mrm{BR}_{e\tau}=\mrm{BR}_{\mu\tau}  =0.
  \end{eqnarray}

Non-vanishing branching ratios to $e\mu$ and $e\tau$ channels are clear indicators for non-zero $\Delta\alpha$.  When $\alpha_1 =\alpha_2=\alpha$ branching ratios to both $e\mu$ and $e\tau$ channels are  very close to zero  $\mrm{BR}_{e\mu}=\mrm{BR}_{e\tau}\approx0$. A small non-zero contribution can be added when $\theta_{13}$ has a value that is close to its upper limit and higher order effects (non-zero $\sin^2{\theta_{13}}$) become influential.
   
Very clearly recognizable signature appears when both Majorana phases are maximal ($\alpha_{1,2}=\pi$). If we also assume $\theta_{23}=\pi/4$ (the best fit value), then $\Phi^{\pm\pm}$ has only two possible decay channels predicting  $\mrm{BR}_{ee}=0.34$ and $\mrm{BR}_{\mu\tau}=0.66$, while all other channels are completely suppressed. Small deviations in $\theta_{23}$ cause small contributions to the $\mu\mu$ and $\tau\tau$ channels while the branching ratio to $\mu\tau$ channel is decreased by the same amount.

\begin{figure}[t]
\begin{center}
\includegraphics[width=0.49\textwidth]{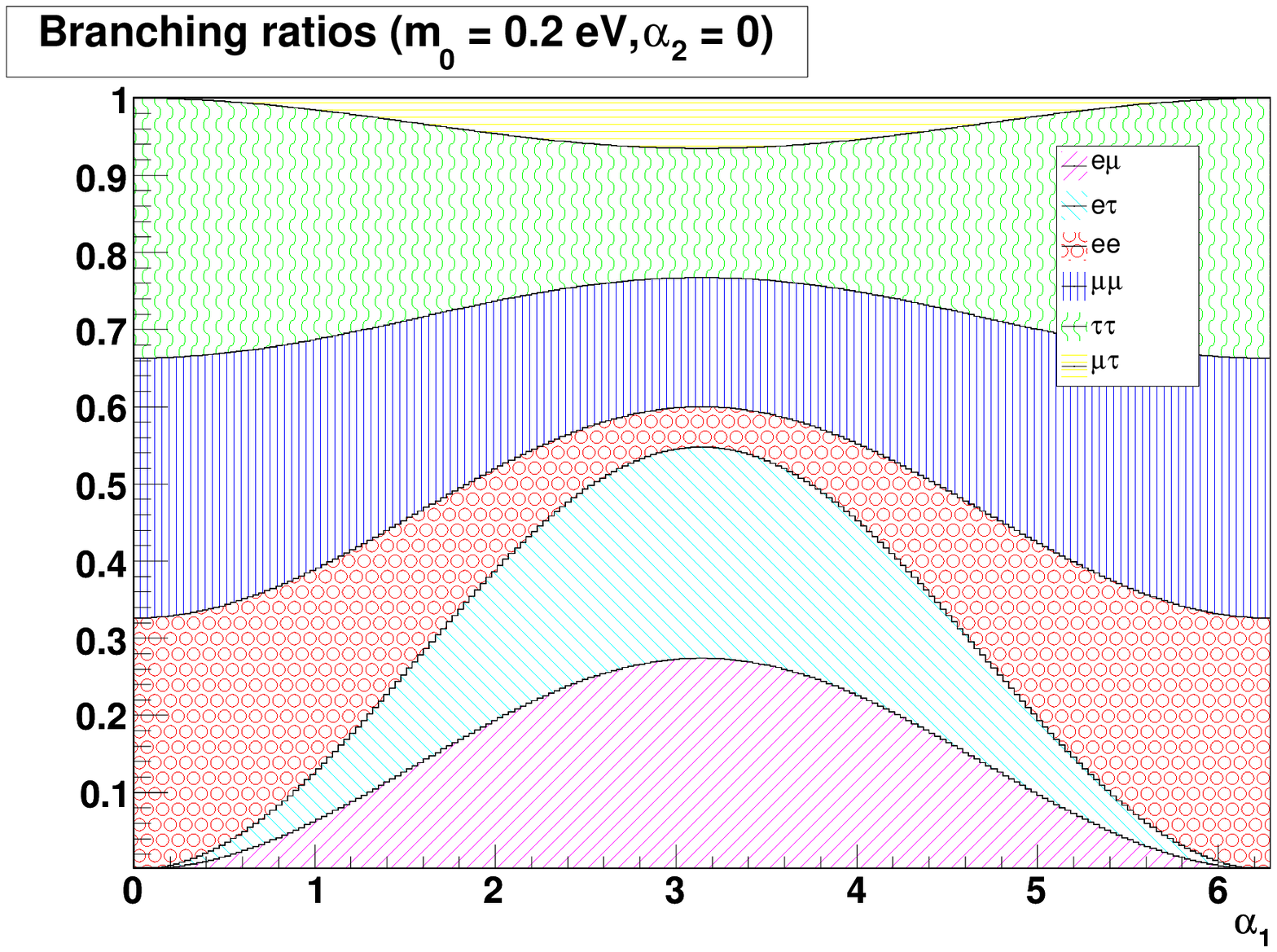}
\hfill
\includegraphics[width=0.49\textwidth]{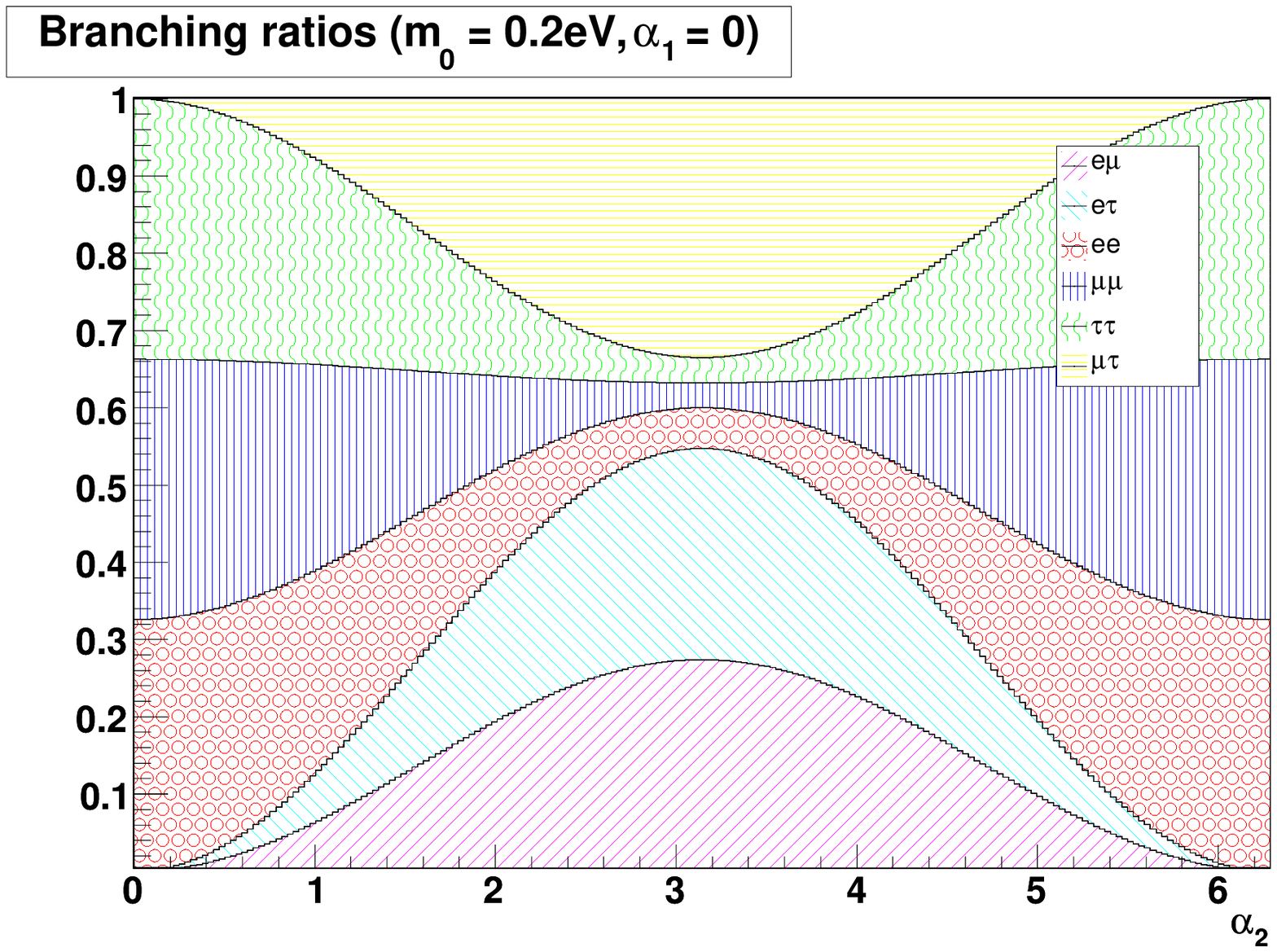}
\caption{Distributions of the $\Phi^{\pm\pm}$  branching ratios for non-zero $\Delta\alpha$.  The left figure presents the case for fixed $\alpha_2=0$, and the right figure for fixed $\alpha_1=0$.}
\label{f3}
\end{center}
\end{figure}

\begin{figure}[h]
\begin{center}
\includegraphics[width=0.49\textwidth]{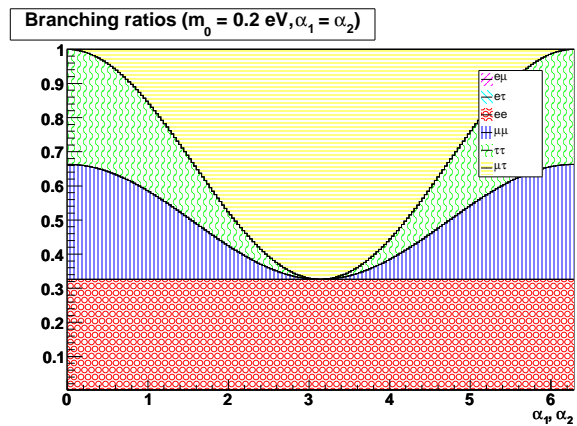}
\caption{Distributions of the $\Phi^{\pm\pm}$  branching ratios for different values of $\alpha_1$ and $\alpha_2$, assuming $\Delta \alpha=0$.}
\label{f2}
\end{center}
\end{figure}

The behavior of branching ratios is plotted in Figure \ref{f3} and Figure \ref{f2} which present the dependence of branching ratios on $\Delta\alpha$. The case for $\alpha_2 = 0$ is shown in the left panel and the one for $\alpha_1 = 0$  in the right panel.  
Figure \ref{f2} shows the branching ratios for different values of Majorana phases when $\Delta\alpha = 0$. 
If we have identified the degeneracy of neutrino masses, we can analyze the values of Majorana phases  without making any assumption about the values of mixing angles.
\begin{itemize}
\item Equal branching ratios to the $ee$, $\mu\mu$ and $\tau\tau$ channels with all other channels being suppressed indicates that $\alpha_1=\alpha_2=0$. 
\item Non-zero branching ratio to the $\mu\tau$ channel means that at least one of the Majorana phases has to be non-zero. 
\item  Non-zero branching ratios to the $e\mu$ and $e\tau$ channels and the deviation
from the result $\mrm{BR}_{ee} = 0.34$ can be generated only by non-zero $\Delta\alpha$. Small non-zero contribution to the $e\mu$ and $e\tau$ channels can be alternatively caused by a 
large  value of $\theta_{13}$.
\end{itemize}


To give exact solutions for the Majorana phases, we fix the values of mixing angles according to the tri-bi-maximal model.
$|\Delta\alpha|$ can be found from Eq.~\eqref{invalpha} which is valid both for the inverted mass hierarchy and the degenerate spectrum.
Separate values for $\alpha_1$ and $\alpha_2$ can be determined from the equation system
\bea
C_5 & \equiv &  \frac{2 \mrm{BR}_{\mu\tau}-\mrm{BR}_{ee}}{2\mrm{BR}_{\mu\mu}+\mrm{BR}_{\mu\tau}+\mrm{BR}_{e\mu}}=
\frac{1}{6}(3-2\cos{\alpha_1}-4\cos{\alpha_2}),
\nonumber
\\
\cos{\Delta\alpha}&= & \cos{\alpha_1}\cos{\alpha_2}+\sin{\alpha_1}\sin{\alpha_2}=
\frac{4-5C_2}{4(1+C_2)}.
\eea
Again, two possible sets of solutions are found for the Majorana phases  and the $\Phi^{\pm\pm}$ branching ratios do not provide the information to decide which of the solutions is correct.

\subsection{Effects of non-zero $\theta_{13}$}
If our assumption about the exact tri-bi-maximal neutrino mixing should not be valid,  the form of previously obtained solutions would also be changed. However, the results and methodology would generally remain the same. Hopefully new upcoming neutrino oscillation experiments will measure the mixing angles with improved precision in the near future \cite{fixtheta}. Small changes in $\theta_{23}$ or $\theta_{12}$ would not affect the structure of the found solutions and we would only need to substitute different values for the mixing angles. Qualitative changes in the analytical expressions appear if $\theta_{13}$ is taken to be non-vanishing. This influences the structure of the solutions and makes the CP-violating angle $\delta$ a physically measurable quantity. 
We note, however, that, due to the smallness of $\sin{\theta_{13}}$, the effect of $\theta_{13}$ and $\delta$ would enter to the branching ratios as a small correction, and extremely precise measurements would be required to detect it. The goal of this section is to analyze the effect of non-zero $\theta_{13}$ to the previously found solutions for Majorana phases and neutrino masses.

In the following  we have still assumed $\theta_{13}$ to be a small parameter
and considered only the leading terms in the expansion with respect to it. We can find the lowest neutrino mass from the similar equation as for $\theta_{13} = 0$ (see Eq.~\eqref{m1y}), only the measured parameter $C_1'$ would differ slightly:
\begin{equation}
C_1^\prime \equiv
\frac{2\mrm{BR}_{\mu\mu}+2\mrm{BR}_{\tau\tau}+2\mrm{BR}_{\mu\tau}-2\mrm{BR}_{ee}}{2\mrm{BR}_{ee}+\mrm{BR}_{e\mu}+\mrm{BR}_{e\tau}} =
\frac{-m_1^2+m_2^2+3m_3^2}{2m_1^2+m_2^2} +{\cal O}(\sin^2\theta_{13}).
\end{equation}
Similarly the determination of Majorana phases has exactly the same structure as earlier. $\Delta{\alpha}$ can be determined uniquely and two possible sets of solutions are found for $\alpha_1$ and $\alpha_2$ when we attempt to determine the absolute values of Majorana phases. 

In conclusion, assuming small but non-zero $\theta_{13}$ does not significantly complicate the determination of interesting neutrino parameters at colliders. The solutions would only be slightly more complex and involve more decay channels as the relations 
$\mrm{BR}_{e\mu} = \mrm{BR}_{e\tau}$ and $\mrm{BR}_{\mu\mu} = \mrm{BR}_{\tau\tau}$ no longer hold. There is a theoretical possibility to find solutions also for $\theta_{13}$ and $\delta$, but such solutions are very sensitive to experimental errors and, in practice, cannot be used at the LHC.

\subsection{Estimation of the impact of experimental uncertainties}

In this section we 
consider  the effects of experimental uncertainties to the determination of $\Phi^{\pm\pm}$ leptonic branching ratios at colliders and, consequently, to the determination of neutrino parameters
in collider experiments. 
The sources of the uncertainties under consideration are.
\begin{itemize}
\item \textbf{Statistical errors} that are relevant for a small number of reconstructed events. 
In this case the number of events observed in particle colliders follow the Poisson statistics with theoretically expected average number of events as a mean value. 
\item \textbf{Random measurement errors} that dominate in the case of large statistical samples and result from the random uncertainties in particle detection and event reconstruction. The measurement errors can vary greatly for different decay channels and their values are strongly experiment and detector specific. 
\item \textbf{Systematical measurement errors} in particle detection and event reconstruction that are negligible for well calibrated detector and correct event reconstruction phenomenology.
\end{itemize}

For the numerical simulation of experimental uncertainties we have firstly modified the theoretically expected number of doubly charged Higgs production events $N^{theor}$ with the Poisson distribution, then calculated and normalized the corresponding branching ratios and finally modified them with Gaussian distortion functions to account for random measurement errors. Possible systematical errors have been neglected. Note that for $\Phi^{\pm\pm}$ pair production each detected event comprises two doubly charged Higgs decays and $N^{theor} = 2 N^{events}$.

In reality the measurement errors are different for different decay channels and their values depend on the specific detector. As full detector-specific error analysis is out of the scope of this paper, we have used uniform uncertainties for all branching ratios for the rough estimation of the effect. In particular, we have assumed Gaussian distortion functions with $\sigma_{BR} = 0.1 \mrm{BR}^{theor}_{ij}$, where $\mrm{BR}^{theor}_{ij}$ is the theoretically expected branching ratio into the corresponding decay channel and $i,j = e, \mu, \tau$. Finally we have run the simulation with randomly distorted branching ratios for $50 000$ times, calculating each time the neutrino parameter of interest. As a result we get the distribution function of particular neutrino parameter which measures the stability of previously found analytical solutions.

\subsubsection{Mass hierarchy determination}
We remind that the neutrino mass hierarchy is identified by the parameter $C_1$, defined in Eq.~\eqref{m1y}, as follows: $C_1>1$ corresponds to the normal hierarchy, $C_1<1$ to the inverted hierarchy and $C_1 \approx1$ to the nearly degenerate mass spectrum. In general, if the lowest neutrino mass is close to zero, the mass hierarchy is very well determined. When the mass increases, the distribution of branching ratios is reaching the nearly degenerate limit and the mass hierarchy or $\mathrm{sign}(\Delta m_{atm})$ is increasingly more difficult to measure.

\begin{figure}[t]
\begin{center}
\includegraphics[width=0.49\textwidth]{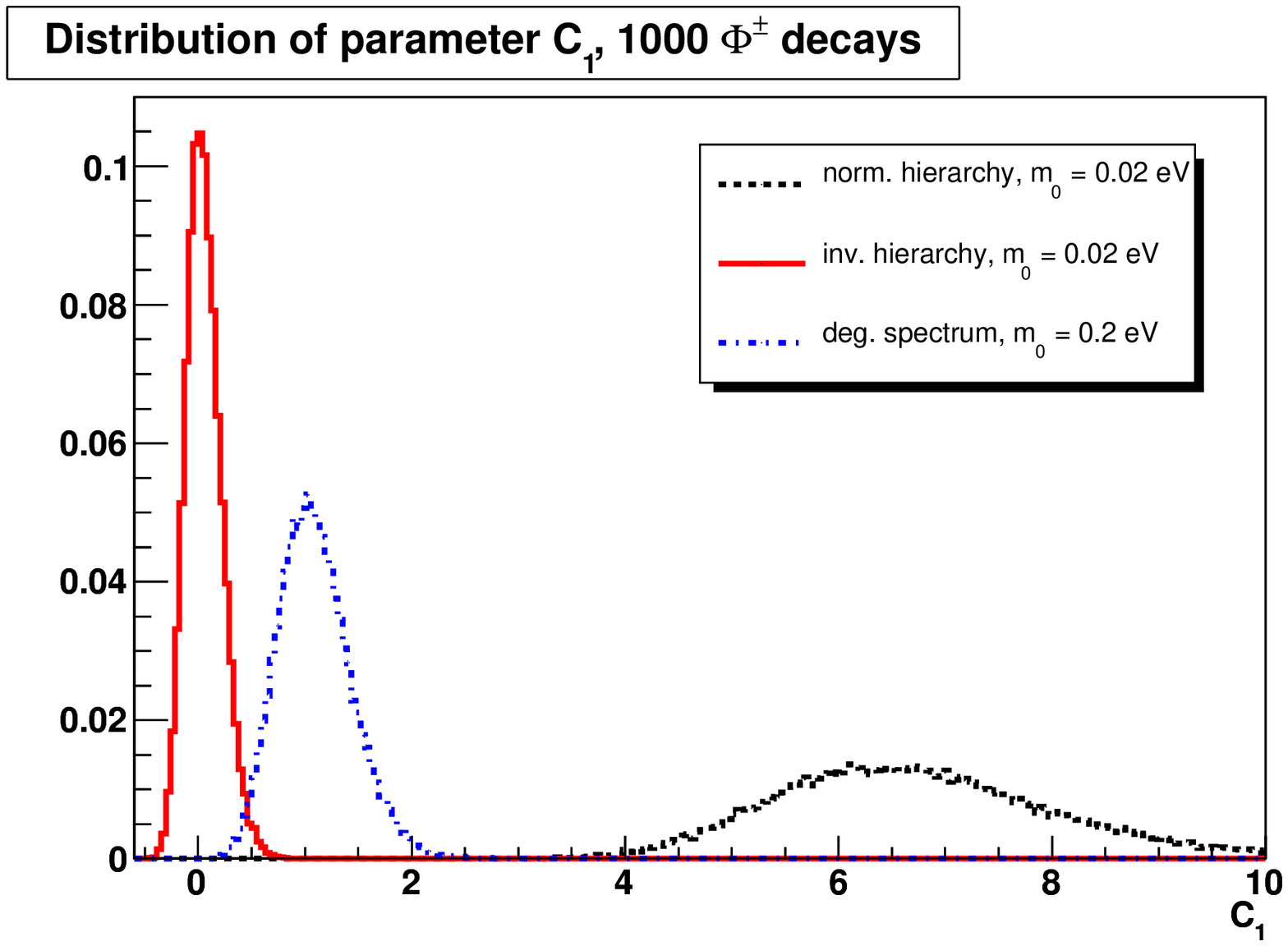}
\hfill
\includegraphics[width=0.49\textwidth]{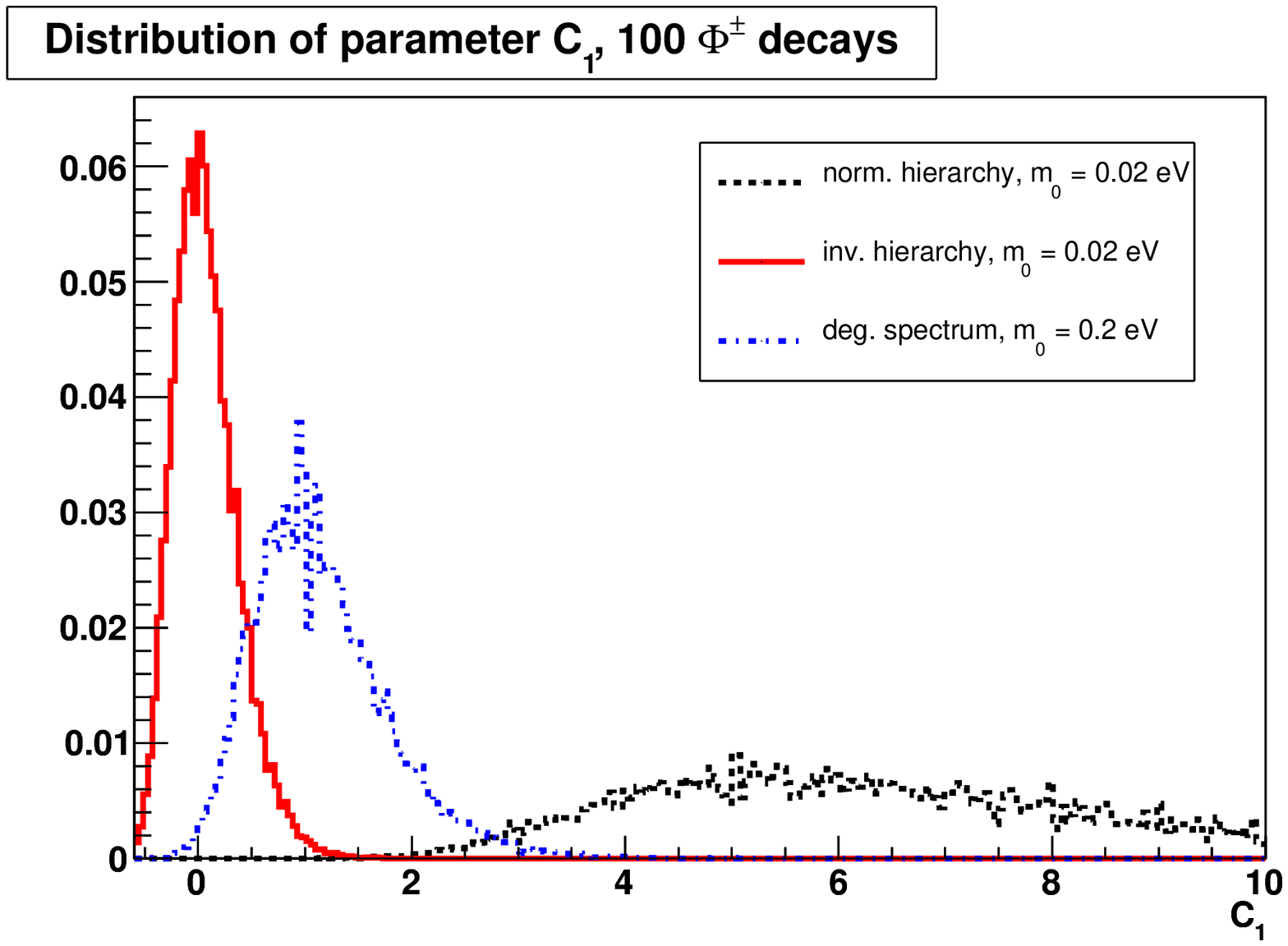}
\caption{Simulated distributions of the parameter $C_1$ due to experimental errors for statistical 
samples of 1000 and 100 $\Phi^{\pm\pm}$ decays in the left and right panel, respectively.
We have assumed   $\sigma_{BR} = 0.1 \mrm{BR}^{theor}_{ij}$ for the branching ratio measurement errors. The solid line represents the inverted hierarchy with $m_0 = 0.02$ eV, dot-dashed line the degenerate spectrum with $m_0 = 0.2$ eV and dashed line the normal hierarchy with $m_0 = 0.02$ eV.}
\label{hierarchy}
\end{center}
\end{figure}

As an example we have analyzed  the behavior of $C_1$ for three different cases: the normal hierarchy for $m_{0} = 0.02$ eV, the inverted hierarchy for $m_{0} = 0.02$ eV and the nearly degenerate limit for $m_0 = 0.2$ eV. The results are presented in  Figure \ref{hierarchy} which shows the 
simulated experimental distribution of $C_1$ for two cases with different statistical samples of events.
Those imply the following $1\sigma$ errors. 

\begin{itemize}
\item Normal hierarchy ($m_0 = 0.02$ eV) 
	\begin{itemize}
		\item 1000 $\Phi^{\pm\pm}$ decays: $C_1 =  6.6 \pm 1.1 \gg 1,$ 
		\item 100 $\Phi^{\pm\pm}$ decays: $C_1 = 6.6 \pm 2.1 \gg 1.$ \\
	\end{itemize}
\item Degenerate limit ($m_0 = 0.2$ eV)
	\begin{itemize}
		\item 1000 $\Phi^{\pm\pm}$ decays: $C_1 =  1.0 \pm 0.3,$ 	
		\item 100 $\Phi^{\pm\pm}$ decays: $C_1 = 1.0 \pm 0.5.$
	\end{itemize}
\item Inverted hierarchy ($m_0 = 0.02$ eV)
	\begin{itemize}
		\item 1000 $\Phi^{\pm\pm}$ decays: $C_1 =  0.06\pm0.15  \ll 1,$ 	
		\item 100 $\Phi^{\pm\pm}$ decays: $C_1 = 0.06\pm 0.28 \ll 1.$
	\end{itemize}
\end{itemize}

The results show that sufficiently good
hierarchy detection accuracy is achieved for small lightest neutrino masses
in both the normal and the inverted hierarchy cases. The accuracy decreases when the mass $m_0$ increases and $C_1 \rightarrow 1$. Such tendency can be understood by comparing the normal and the inverted hierarchy plots in Figure \ref{mass} where the distribution of branching ratios clearly differs for small mass values and becomes very similar to each other when the mass is increased. Figure \ref{hierarchy} shows that the normal hierarchy is very well determined by the parameter $C_1$ even for small statistics while at $3\sigma$ level
  the inverted hierarchy can be confused with the degenerate mass
spectrum for small statistics. The main factor that clearly distinguishes the normal mass hierarchy with small $m_0$ is the negligible value of $\mrm{BR}_{ee}$ that can not occur for the inverted or degenerate spectra.

\subsubsection{$m_0$ measurement}
After the neutrino mass hierarchy has been determined, we can use either Eq.\eqref{norm} (for the normal hierarchy) or Eq.\eqref{invmass} (for the inverted hierarchy) to estimate the value of the lowest neutrino mass. In the following we analyze the achievable precision for these parameters. Again, for small values of $m_0$ the measurement precision is sufficiently high and decreases when
$m_0$ approaches the degenerate values. 

As already verified by the hierarchy determination accuracy, the normal hierarchy provides a distinguishable signature and could thus be easily identified, while the inverted hierarchy can be confused with the degenerate spectrum. Such a tendency is also notable in the measurement of lowest neutrino mass. For the normal hierarchy not only hierarchy  but also the actual value of the lowest neutrino mass can be measured with relatively good precision. Turning back to our earlier example we assume the true value of the lowest neutrino mass to be $m_1 = 0.02$ eV,  $\sigma_{BR} = 0.1 \mrm{BR}^{theor}_{ij}$ to be the branching ratio measurement errors at collider experiments, and find that the $1\sigma$ experimental errors for the lightest neutrino masses are $\delta m_1 =2\cdot 10^ {-3}$ eV and $\delta m_1 = 5\cdot 10^{-3}$ eV for the statistical samples of 1000 and 100 $\Phi^{\pm\pm}$ decays, respectively. In order to measure $m_3$ from Eq.~\eqref{invmass} for the inverted mass hierarchy with comparable precision, a very good statistical basis (more than 5000 events) and the measurement errors smaller than $\sigma_{BR} = 0.01 \mrm{BR}_{ij}^{theor}$  are required. The data of such quality would not be obtainable from the LHC  experiments but only from the future colliders (possibly ILC). 

\subsubsection{Measurement of Majorana phases}
Determination of Majorana phases is discussed in detail in Section \ref{appr}. It  depends on the neutrino mass hierarchy. For the normal mass hierarchy with small $m_1$ it is very difficult to estimate the Majorana phases with realistic measurement errors. This is due to the fact that the only observable $\alpha_2$ does not significantly influence the distribution of branching ratios (see Figure \ref{fn}).
To the contrary, for the degenerate spectrum or inverted hierarchy  the Majorana phases strongly influence the distribution of branching ratios which, in principle, can be measured in realistic experimental conditions. As an example we estimate the measurement error for $\Delta\alpha$  for the inverted hierarchy. As we have shown earlier in Eq.~\eqref{invalpha}, to high accuracy such a calculation does not depend on the value of $m_3$. We find that  the $1\sigma$ errors for $\Delta\alpha$ are $0.06 \pi $ and $0.03\pi$ for 100 and 1000 $\Phi^{\pm\pm}$ decays, respectively. Similar precision is achieved assuming the degenerate spectrum. This result is general and does not depend considerably on the particular value of $\Delta \alpha$.

Full detector-specific analysis for the measurement errors of branching ratios requires separate analyzes. The error estimations that are found in this section are only approximate, but still emphasize the promising nature of our method for determining neutrino parameters in particle collider experiments.

\section{Determination of triplet Higgs VEV} \label{Sec4}

In our scenario the neutrino mass matrix is directly related to the doubly charged Higgs
leptonic branching fractions  according to \Eq{mnu}, and
 the overall normalization factor is the triplet Higgs VEV $v_{\Phi}$.
Therefore, one needs additional experimental measurements for determination
of $v_{\Phi}$, and thus the entire low energy neutrino mass matrix. Those measurements can come 
either from collider physics, from the low energy neutrino mass measurements or from cosmology.

Let us first assume that $v_{\Phi}$ is large enough to imply, according to  Eq.~\eqref{dwW},
observable fraction of the decays $\Phi^{++}\to W^+W^+$, and the collider experiments are sensitive
enough to measure not just the branching fractions but also the partial widths of the triplet, 
namely $\Gamma_{ij}$ and $\Gamma_{WW}$. 
The latter may not be possible at LHC but could be possible at ILC experiments \cite{Barenboim:1996pt}.
In such a case one gets from  Eq.~\eqref{dw} and  Eq.~\eqref{dwW}, 
\begin{equation}
\frac{\mrm{BR}_{ll}}{\mrm{BR}_{WW}} = \frac{\Gamma_{ll}}{\Gamma_{WW}}=\frac{\Gamma_{ll}}{k v_{\Phi}^2}\Rightarrow v_\Phi = \sqrt\frac{\Gamma_{ll}\mrm{BR}_{WW}}{k \mrm{BR}_{ll}}=\sqrt{\frac{\Gamma_{tot} \mrm{BR}_{WW} }{k}},
\end{equation}
and the determination of $v_{\Phi}$ from collider experiments is possible.

If the collider experiments are not able to measure the partial widths of the triplet, one 
needs additional information on the neutrino mass matrix. Assuming that the 
branching ratio to $WW$ channel is measured at any accelerator experiment and $|(m_\nu)_{ee}|$ is probed from $0\nu\beta\beta$ experiment one gets
\begin{equation}
\frac{\mrm{BR}_{ee}}{\mrm{BR}_{WW}} = \frac{\Gamma_{ee}}{\Gamma_{WW}} = \frac{1}{32\pi}\frac{|(m_\nu)_{ee}|^2 m_{\Phi^{\pm\pm}}}{k v_\Phi^4}.
\end{equation}
Now $v_{\Phi}$ can be directly found as
\begin{equation}
v_{\Phi} = \left(\frac{|(m_\nu)_{ee}|^2 m_{\Phi^{\pm\pm}} \mrm{BR}_{WW}}{32\pi k \mrm{BR}_{ee}}\right)^\frac{1}{4}.
\end{equation}

Finally, if $v_{\Phi}$ is too small to imply observable  $\Phi\to WW$ decay rates at colliders, 
one has to 
rely entirely on leptonic data. 
If one of the leptonic Yukawa couplings is directly measured in the accelerator experiments and $|(m_\nu)_{ee}|$ is probed from the $0\nu\beta\beta$ experiments, one is able to derive the VEV from data.
As the simplest example, when $\Gamma_{ee}$ is measured, perhaps from the resonance at $e^-e^-$
collider \cite{Raidal:1997tb}, the VEV can be directly found from
\begin{equation}
\Gamma_{ee} = \frac{|(m_\nu)_{ee}|^2m_{\Phi^{\pm\pm}}}{32\pi v_{\Phi}^2} \Rightarrow v_\Phi = \sqrt{\frac{|(m_\nu)_{ee}|^2 m_{\Phi^{\pm\pm}}}{32\pi\Gamma_{ee}} }.
\end{equation}

As shown, direct measurement of the VEV is possible. However it does require additional information which cannot be obtained from the LHC alone. Should the $0\nu\beta\beta$ yield positive results or
some of the triplet Yukawa coupling be measured at  ILC,  we can also give estimates on 
the magnitude of the VEV of the Higgs triplet.

\section{Conclusions} 
\label{Sec5}

The main motivation for the present paper is to study how to test the TeV scale triplet Higgs neutrino
mass mechanism directly at collider experiments. From the collider physics point of
view this mechanism has several advantages  over the singlet neutrino
mass mechanism. As the triplet has gauge quantum numbers, its production at colliders 
is limited only by the mass reach not by tiny Yukawa couplings as is the case for singlets. 
Thus several hundreds of those particles can be produced at LHC and ILC experiments. 

The branching ratios of doubly charged Higgs decays to two same charged leptons 
directly probe the corresponding element of  the neutrino mass matrix. This allows
us to study what one can learn about the light neutrino parameters from collider 
experiments. We have shown that  the neutrino mass ordering, the lightest neutrino mass and 
the Majorana phases can be measured at colliders  by just  counting the lepton flavours. 
We emphasize that those are exactly  these neutrino parameters which present neutrino
oscillation experiments are not sensitive to. Therefore collider tests of neutrino mass
mechanism may provide a major breakthrough in neutrino physics. 

We find that there are some flavour combinations of the doubly charged Higgs decay products
which definitely point towards certain solutions.
For example, should LHC see only doubly charged Higgs decays to muons and taus, light neutrinos
must have strong normal hierarchy, and the lightest neutrino mass can be measured. 
In particular, the observation or non-observation of $ee$ final states is a clear
discriminator between the mass hierarchies.   
Similarly, in the optimistic scenarios discussed in Section \ref{Sec3},
one can estimate the magnitude of the Majorana phase(s) of light neutrinos. In less 
clear cases, however,
the experimental errors of the collider experiments may jeopardize  
the neutrino parameter measurements and no definite conclusion  can be drawn.

We have also shown that one can actually fully determine the light neutrino mass matrix
from collider experiments and/or from the measurement of neutrinoless double
beta decay parameters. 
This requires determination of the triplet Higgs partial widths to leptons and to gauge bosons
which could be possible at ILC experiments. If the triplet Higgs turns out to be light enough to be produced at colliders, neutrino physics may get an unexpected contribution from 
collider experiments.

\vskip 0.5in

\section*{Note added} 
\noindent  When the research presented in this paper was completed, an
e-print \cite{Garayoa:2007fw} appeared in arXives addressing the same topic. 
Our numerical results are in agreement with theirs. 
However, our results on neutrino parameters are also obtained  in an analytical form 
which is not the case in  \cite{Garayoa:2007fw}.

\vskip 0.5in
\vbox{
\noindent{ {\bf Acknowledgments} } \\
\noindent  
We thank A. Hektor and M. M\"untel for discussions and J. Garayoa for pointing out 
an error in the first version of this paper. 
This work is partially supported 
by ESF grant No 6140, by EC I3 contract No 026715 and
by the Estonian Ministry of Education and Research.   

}

\end{document}